# Andreev Reflection Enhancement in Semiconductor-Superconductor Structures

Shlomi Bouscher[1], Roni Winik[1,2] and Alex Hayat[1]

[1] *Department of Electrical Engineering, Technion, Haifa 32000, Israel*
[2] *Solid State Institute, Technion, Haifa, 32000, Israel*

We develop a new theoretical approach for modelling a wide range of semiconductor-superconductor structures with arbitrary potential barriers and a spatially-dependent superconducting order parameter. We demonstrate asymmetry in the conductance spectrum as a result of a Schottky barrier shape. We further show that Andreev reflection process can be significantly enhanced through resonant tunneling with appropriate barrier configuration, which can incorporate the Schottky barrier as a contributing component of the device. Moreover, we show that resonant tunneling can be achieved in superlattice structures as well. These theoretically demonstrated effects along with our modelling approach enable much more efficient Cooper pair injection into semiconductor-superconductor structures, including superconducting optoelectronic devices.

# I. INTRODUCTION

Semiconductor-superconductor hybrid devices are a growing field of research, which holds promising applications due to the superconducting proximity effect and the ability to use various nontrivial quasiparticle states for quantum devices and applications. Many types of hybrid semiconductor-superconductor devices have been successfully fabricated [1,2,3,4] and characterized, with some of the experimental results showing interesting phenomena, such as asymmetry in the conductance spectrum [1]. Moreover, various applications involving semiconductor-superconductor structures have been shown [5,6,7,8,9,10,11]. These include superconducting light emitting diodes (SLEDs) [5,6,7,9] and superconducting quantum dots capable of emitting entangled photon-pairs [8], Bell-state analyzers [10] and semiconductor-superconductor based waveguide amplifiers [11]. This wide range of novel devices takes advantage of Andreev reflection [12] at the superconductor-semiconductor interface [1,2,3,4], to directly inject electron Cooper-pairs into the semiconducting side of the junction. The injected electron pairs can then undergo radiative recombination with holes to generate entangled-photon pairs, which are of great importance to the field of quantum information [13,14]. The Cooper-pair injection efficiency is highly dependent on the properties of the materials used, such as the coherence length of the superconductor and the mean free path in the normal material. Moreover, the shape of the potential barrier at the junction, chemical potentials and effective masses can have a considerable effect on carrier transport.

A simple but very successful theory for describing the transport mechanisms in superconductor-normal (SN) junctions is the Blonder-Tinkham-Klapwijk (BTK) model [15] which assumes a delta-like potential barrier and provides an analytical approach for calculating the transport properties of the SN barrier based on the Bogoliubov-de Gennes (BdG) equations. The BTK model has been applied to describe various SN junction types, including superconductor-magnetic semiconductor junctions [16], superconductor-ferromagnet junctions [17,18] and metal-heavy-fermion superconductor junctions [19,20]. Moreover, the BTK model has been modified to include angle dependence [21] and high-$T_c$ superconductors [22]. However, these approaches are based on the delta-barrier approximation, and in order to obtain results for structures involving elaborate potential barriers, full calculation of coupled BdG equations must be performed using the properties and the band structure of the semiconductor.

Here we develop a complete model based on the finite element approach, enabling analysis of a wide range of semiconductor-superconductor structures with arbitrary potential barriers and a spatially dependent superconducting order parameter. Furthermore, we show that by designing the spatial dependence of semiconductor bandgap, the Andreev reflection process can be enhanced by resonant tunneling with an appropriate semiconductor structure. The enhancement can be achieved using a single quantum well, or with a superlattice design. We calculate the Cooper pair injection efficiency and the reflection/transmission efficiencies while taking into account different effective masses and Fermi levels of each of the materials comprising the device. Results obtained match the BTK model for the simple delta-barrier approximation, and display new behavior for enhanced Andreev-reflection structures.

## II. FINITE ELEMENT MODEL

Our model is based on the Bogoliubov-de Gennes (BdG) equations resulting from the Bardeen-Cooper-Schrieffer (BCS) model [23]. The BdG equations describe the behavior of superconductors at finite temperatures with electron-hole quasi-particles which are formed as excitations of the BCS ground state and therefore depend on the binding term $\Delta(x)$ - the coupling parameter of the superconductor. The BdG equations can be written in the following form:

$$[-\frac{\hbar^2}{2}\nabla\frac{1}{m(x)}\nabla - \mu(x) + V(x)]u(x,t) + \Delta(x)v(x,t) = i\hbar\frac{\partial u(x,t)}{\partial t}$$
$$-[-\frac{\hbar^2}{2}\nabla\frac{1}{m(x)}\nabla - \mu(x) + V(x)]v(x,t) + \Delta(x)u(x,t) = i\hbar\frac{\partial v(x,t)}{\partial t}$$

(1)

where $u(x,t), v(x,t)$ represent the quasi-particle wavefunctions, $V(x)$ is an arbitrary spatial potential and $\mu(x)$ is the chemical potential, which at low temperatures of superconductivity can be approximated to be equal to $E_f$ - the Fermi energy. In our model, $V(x)$ is taken to be equal to $E_c(x)$ [24,25]. Since $E_c(x)$ is defined as the bottom of the conduction band, the kinetic energy of the quasiparticles is $E_k(x) = E_{tot}(x) - E_c(x) = E_{tot}(x) - V(x)$. The combination of different materials in our model requires including a spatially variant effective mass $m(x)$ (Eq. 1). Spatially varying parameters have been shown to result in the Hamiltonian becoming non-Hermitian [26], preventing the use of the standard kinetic term $\frac{\hbar^2}{2m(x)}\nabla^2$ in our model. The correct Hermitian form of the kinetic term has been proven to be $\frac{\hbar^2}{2}\nabla\frac{1}{m(x)}\nabla$ [27]. At the core of the finite element method (FEM) approach used in our modeling is the division of a given domain into $N$ equal segments along the $x$ axis (perpendicular to the SN interface), where the solution of the entire problem can be represented by a sum of characteristic basis functions $\phi_n$ spanning each $n$-th segment [28,29]:

$$\Psi = \sum_n a_n \phi_n \quad (2)$$

When solving the Schrodinger equation in its stationary form, the characteristic basis functions $\phi_n$ can be chosen to be the solutions of the Schrodinger equation. In our model, the segments were assumed to be small enough so that the potential in each segment is approximately constant (piecewise constant approximation). This allows choosing plane waves as the basis functions for our total solution as they are the exact solutions for each segment [30]. In addition, we further divide the structure into two regions: the barrier region which includes the arbitrary potential and has $N_1$ segments, and a termination region including $N_2$ segments with $N_1+N_2=N$ (Fig 1 b,c). The additional termination region is defined to ensure proper scattering formulation and is located inside the superconductor. Thus, for the steady-state case, the solutions for the BdG equations for a specific segment are coupled plane waves with the following energy momentum relations:

$$k_n^{\pm}(E_n) = \frac{\sqrt{2m_n}}{\hbar} \left[ E_f - V_n \pm \left( E_n^2 - \Delta_n^2 \right)^{\frac{1}{2}} \right]^{\frac{1}{2}} \quad (3)$$

where $k_n^+$, $k_n^-$ denote the momenta of the quasiparticles involved, $V_n$, $\Delta_n$, $m_n$, $E_n$, $E_f$ are the values of the potential barrier, the coupling coefficient, effective mass, quasiparticle energy and the Fermi energy for the *n*-th segment respectively (Fig. 1). As our model and calculations extend to a relatively small energy range around the superconducting gap parameter $\Delta$, they satisfy the requirement of the long-wavelength or low-excitation limit of BdG theory. In general, arbitrary potential barriers $V(x)$ do not have to be continuous. Therefore, even for discontinuities in the effective mass, potential or superconducting gap, the basic requirement for the continuity of the wavefunction is always fulfilled.

Since the arbitrary potential in our model is approximated in terms of small segments, in which the barrier is constant, the only possible change between the potentials of neighboring segments is a step function (Fig 1c). These step functions are *artificial* discontinuities resulting from the finite resolution of the finite-element model – in contrast to the *physical* discontinuities which are an inherent part of the problem. Thus, the requirement of a large number of segments $N$ is necessary in order to obtain an accurate numerical solution to satisfy the piecewise constant approximation. It has been shown [31] that for the one-dimensional stationary Schrodinger equation, the error of the solution scales as the inverse of the number of segments $N$ for a piecewise constant potential approximation. Therefore, increasing the number of segments strongly enhances the accuracy of the resulting solution. Moreover, as the number of segments $N$ increases, the artificial discontinuities resulting from the numerical model that occur between the neighboring segments vanish, leaving only the original physical discontinuities which must be accounted for in any solution – numerical or analytical. For the case of effective mass spatial variation described by a step function, it was shown that the Hermitian form of the Hamiltonian (Eq. 1) leads to effective-mass-dependent boundary conditions [27]:

$$\begin{cases} \psi_1 = \psi_2 \\ \dfrac{1}{m_1}\dfrac{\partial \psi_1}{\partial x} = \dfrac{1}{m_2}\dfrac{\partial \psi_2}{\partial x} \end{cases} \quad (4)$$

where $m_i$ and $\psi_i$ are the effective mass and wavefunction on each side of the boundary respectively. In our model, the boundary conditions specified above are included in each transfer matrix and thus lead to a proper description and solution of the problem at hand.

The quasi-particle wavefunction in its most general form (including its derivative) for a single segment is:

$$\begin{bmatrix} \Psi_{n_u}(x) \\ \Psi_{n_v}(x) \\ \frac{\partial \Psi_{n_u}(x)}{\partial x} \\ \frac{\partial \Psi_{n_v}(x)}{\partial x} \end{bmatrix} \triangleq \begin{bmatrix} u_n(x) \\ v_n(x) \\ \frac{\partial u_n(x)}{\partial x} \\ \frac{\partial v_n(x)}{\partial x} \end{bmatrix} = A_n \begin{bmatrix} u_n \\ v_n \\ ik_n^+ u_n \\ ik_n^+ v_n \end{bmatrix} e^{ik_n^+ x} + B_n \begin{bmatrix} u_n \\ v_n \\ -ik_n^+ u_n \\ -ik_n^+ v_n \end{bmatrix} e^{-ik_n^+ x} + C_n \begin{bmatrix} v_n \\ u_n \\ ik_n^- v_n \\ ik_n^- u_n \end{bmatrix} e^{ik_n^- x} + D_n \begin{bmatrix} v_n \\ u_n \\ -ik_n^- v_n \\ -ik_n^- u_n \end{bmatrix} e^{-ik_n^- x} \quad (5)$$

where $\Psi_{n_u}(x), \Psi_{n_v}(x)$ describe the two components of the quasi-particle wavefunctions $u(x), v(x)$. The general wavefunction is composed of four different plane waves, each corresponding to a quasi-particle state (electron-like or hole-like) and a propagation direction. For a segment with a given thickness $l_n$, the wavefunctions on each side of the segment are related through the segment scattering, or transfer matrix $M_n$:

$$\vec{\Psi}_n(x = L + l_n) = M_n \vec{\Psi}_n(x = L) \quad (6)$$

where $\vec{\Psi}_n$ includes the quasi-particle wavefunctions and their derivatives, a total of four terms and $L$ is the location of the n-th segment.

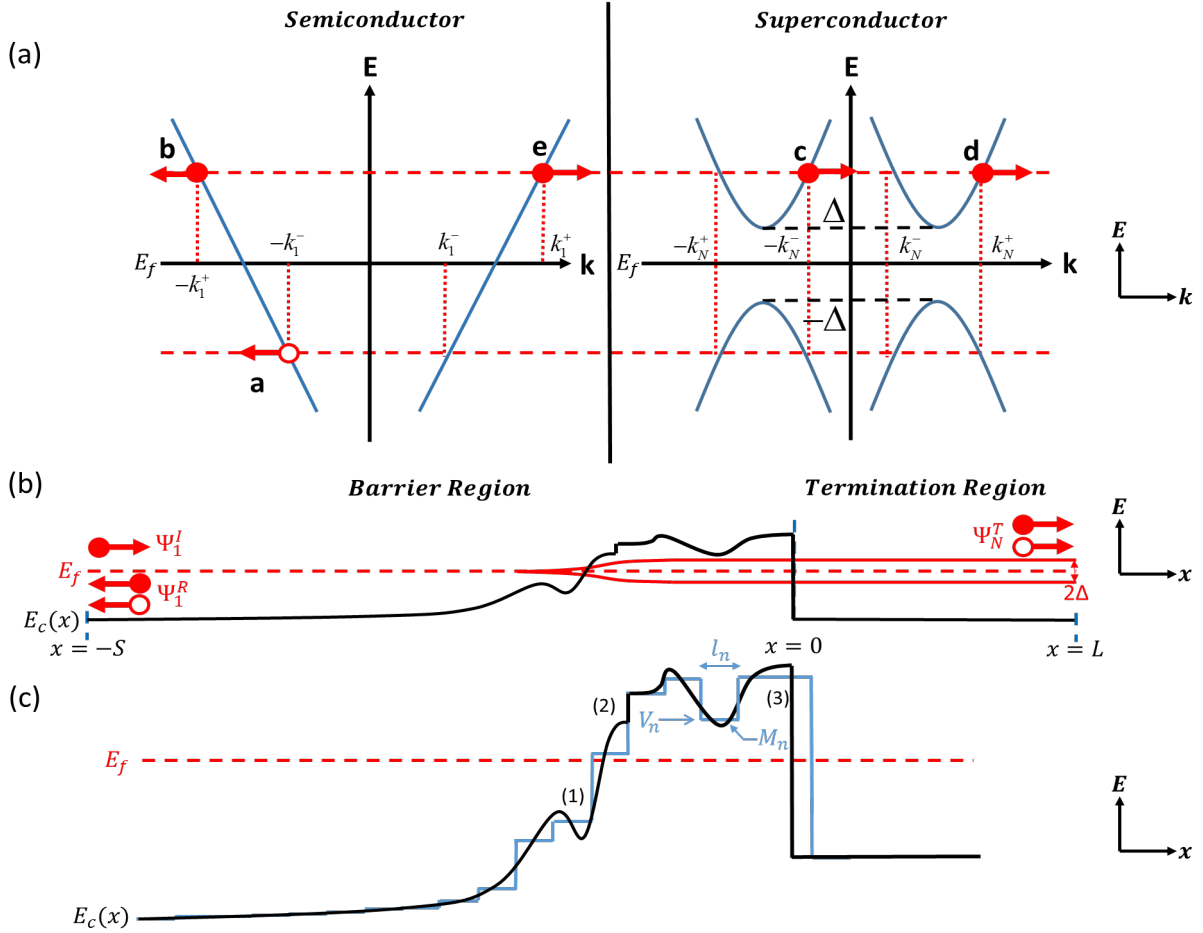

**Fig 1. (a)** A schematic diagram for the dispersion relation on each side of the barrier. **a** represents the reflected hole (Andreev reflection), **b** represents a reflected electron **c** and **d** represent the transmitted quasi-particles and **e** represents an incoming electron. **(b)** A spatial diagram depicting an arbitrary barrier structure as well as the division into two distinct regions; the barrier region and the termination region. The barrier region includes the arbitrary barrier structure along with superconducting gap variation and takes into account both inbound and outbound particles. The termination region includes constant potential and constant supercodncuting gap values, and only supports outbound quasiparticles. **(c)** Closeup of the arbitrary barrier shown in (b). The barrier structure is divided into $N_1$ segments, each with a width $l_n$, constant potential $V_n$ (blue) and a characteristic transfer matrix $M_n$. The same division is performed in the termination region with a total number of $N_2$ segments. Several examples of resolution related issues: (1) demonstrates how a low resolution can remove fine features of the potential barriers such as the small dip. (2) demonstrates that both physical and numerical discontinuities can overlap. (3) demonstrates that low resolution can cause the shift of a physical discontinuity. All of the examples above can be corrected through the use of higher resolution leaving only physical discontinuities.

The general form of the transfer matrix is:

$$M_n = \begin{pmatrix} \gamma_n c_n^+ - \varepsilon_n c_n^- & -\alpha_n c_n^+ + \alpha_n c_n^- & \frac{\gamma_n}{k_n^+}s_n^+ - \frac{\varepsilon_n}{k_n^-}s_n^- & -\frac{\alpha_n}{k_n^+}s_n^+ + \frac{\alpha_n}{k_n^-}s_n^- \\ \alpha_n c_n^+ - \alpha_n c_n^- & -\varepsilon_n c_n^+ + \gamma_n c_n^- & \frac{\alpha_n}{k_n^+}s_n^+ - \frac{\alpha_n}{k_n^-}s_n^- & -\frac{\varepsilon_n}{k_n^+}s_n^+ + \frac{\gamma_n}{k_n^-}s_n^- \\ \frac{m_{n+1}}{m_n}\cdot[-\gamma_n k_n^+ s_n^+ - \varepsilon_n k_n^- s_n^-] & \frac{m_{n+1}}{m_n}\cdot[\alpha_n k_n^+ s_n^+ - \alpha_n k_n^- s_n^-] & \frac{m_{n+1}}{m_n}\cdot[\gamma_n c_n^+ - \varepsilon_n c_n^-] & \frac{m_{n+1}}{m_n}\cdot[-\alpha_n c_n^+ + \alpha_n c_n^-] \\ \frac{m_{n+1}}{m_n}\cdot[-\alpha_n k_n^+ s_n^+ + \alpha_n s_n^-] & \frac{m_{n+1}}{m_n}\cdot[\varepsilon_n k_n^+ s_n^+ - \gamma_n k_n^- s_n^-] & \frac{m_{n+1}}{m_n}\cdot[\alpha_n c_n^+ - \alpha_n c_n^-] & \frac{m_{n+1}}{m_n}\cdot[-\varepsilon_n c_n^+ + \gamma_n c_n^-] \end{pmatrix} \quad (7)$$

with $\alpha_n \triangleq \frac{u_n v_n}{u_n^2 - v_n^2}, \gamma_n \triangleq \frac{u_n^2}{u_n^2 - v_n^2}, \varepsilon_n \triangleq \frac{v_n^2}{u_n^2 - v_n^2}, s_n^\pm \triangleq \sin(k_n^\pm l_n), c_n^\pm \triangleq \cos(k_n^\pm l_n)$, where the coefficients $u_n, v_n$ are the local wavefunction amplitudes, $l_n$ is the width of the segment, $k_n^\pm$ are the wavenumbers an $m_n, m_{n+1}$ are the effective masses of the $n$-th and $n+1$-th segments. These matrices thus fully characterize the physics behind the model presented in our manuscript with similar methods being successfully applied for the case of superlattices [39,32] and arbitrary potentials [33] which also contain physical potential discontinuities and can simulate the effects of delta barriers. Since spatially varying parameters such as the effective mass, superconducting gap parameter and the potential are not necessarily continuous, adjacent transfer matrices $M_n$ and $M_{n+1}$ can be quite different.

Since there are $N$ total segments to the barrier, the transfer matrix for the arbitrary potential barrier $M_{Bar}$ is:

$$\vec{\Psi}_{N_1} = \left(\prod_{n=1}^{N_1-1} M_n\right)\vec{\Psi}_1 = M_{Bar}\vec{\Psi}_1 \quad (8)$$

The termination region has a simplified scattering matrix $M_{Ter}$ with only the outgoing terms, so that the total transfer matrix $M_{Tot}$ is:

$$\vec{\Psi}_N = M_{Ter} M_{Bar} \vec{\Psi}_1 = M_{Tot} \vec{\Psi}_1 \qquad (9)$$

The exterior boundary conditions are:

$$\vec{\Psi}_1^I = \begin{bmatrix} 1 \\ 0 \\ ik_1^+ \\ 0 \end{bmatrix} e^{ik_1^+(-S)}$$

$$\vec{\Psi}_1^R = a \begin{bmatrix} 0 \\ 1 \\ 0 \\ -ik_1^- \end{bmatrix} e^{-ik_1^-(-S)} + b \begin{bmatrix} 1 \\ 0 \\ ik_1^+ \\ 0 \end{bmatrix} e^{ik_1^+(-S)}, \quad \vec{\Psi}_N^T = M_{Tot}(\vec{\Psi}_1^I + \vec{\Psi}_1^R) \qquad (10)$$

$$\vec{\Psi}_N^T = c \begin{bmatrix} u_N \\ v_N \\ ik_N^+ u_N \\ ik_N^+ v_N \end{bmatrix} e^{ik_N^+ L} + d \begin{bmatrix} v_N \\ u_N \\ -ik_N^- v_N \\ -ik_N^- u_N \end{bmatrix} e^{-ik_N^- L}$$

where $-S$ denotes the starting coordinate of the arbitrary spatial potential and $L$ denotes its end. $a, b, c, d$ are the scattering amplitudes, $\vec{\Psi}_1^I$ is the incident electron wavefunction in the semiconductor, $\vec{\Psi}_1^R$ is the reflected electron and hole wavefunction in the semiconductor and $\vec{\Psi}_N^T$ is the transmitted quasiparticle wavefunction in the superconductor. The '1' and 'N' subscripts denote the segments each wavefunction belongs to. Eq. 10 describes the relation between the boundary conditions on either end of the barrier potential, from which scattering amplitudes can be extracted to obtain the reflection/transmission probability coefficients:

$$A(E) = |a|^2 \frac{k_1^-}{k_1^+}$$

$$B(E) = |b|^2$$

$$C(E) = |c|^2 \left(|u_N|^2 - |v_N|^2\right) \frac{m_1}{m_N} \frac{k_N^+}{k_1^+} \tag{11}$$

$$D(E) = |d|^2 \left(|u_N|^2 - |v_N|^2\right) \frac{m_1}{m_N} \frac{k_N^-}{k_1^+}$$

where $A(E)$ is the probability of a hole reflecting (Andreev reflection), $B(E)$ is the probability of an electron reflecting, $C(E), D(E)$ are the probabilities for the transmission of the quasi-particles and $m_1$, $m_N$ are the effective masses of the quasiparticles in the semiconductor and superconductor respectively.

Since current is conserved, it can be calculated at any point along the structure. Similar to the case of the BTK model, the choice in our model is to calculate the current on the normal side of the structure. Thus, an expression for the differential conductance of the junction, which takes both the forward and backward currents into account, can be obtained [15]:

$$\frac{\partial I_{Junction}}{\partial V_{App}} = 2N(0) e v_F A_{Junc} \int_{-\infty}^{+\infty} \left[\frac{df_0(E - eV_{App})}{dV_{App}}\right] [1 + A(E) - B(E)] dE \tag{12}$$

where $f_0$ is the Fermi-Dirac distribution, $V_{App}$ is the applied voltage, $N(0)$, $v_F$, $A_{junc}$ are the normal-material electron density of states (DOS), Fermi velocity and the area of the junction cross section respectively at the cross-section in which the current is being calculated. The calculation of the conductance spectra includes only the derivative of the Fermi-Dirac distribution $df_0(E - eV_{App})/dV_{App}$ rather than the distribution itself. Therefore, while the Fermi-Dirac distribution can vary due to the spatial variation of the band edge, its derivative around the Fermi level, determined by the temperature, will be the same function of energy in different segments – independent of the Fermi level distance from the bottom of the band. For $T=0$, the Fermi-Dirac distributions become

Heaviside step functions with the resulting $f_0(E - eV_{App}) - f_0(E)$ becoming a rectangular window function with a width $eV_{App}$. As a result, Eq. (12) can be further simplified:

$$\frac{\partial I_{Junction}}{\partial V_{App}} \approx 2N(0)e^2 v_F A_{Junc} e \left[1 + A(eV_{App}) - B(eV_{App})\right] \propto 1 + A(eV_{App}) - B(eV_{App}) \quad (13)$$

The conduction spectrum is thus directly proportional to the hole ($A$) and electron ($B$) reflection probabilities. Since the conductance spectrum is typically normalized by the conductance spectrum above $T_c$, coefficients such as $N(0)$, $v_F$ and $A_{Junc}$ are canceled out. Nevertheless, phenomena such as Fermi velocity mismatch at the junction are still taken into account in our model since both the effective mass and $k$-vectors are included in the transfer matrices. Therefore, any mismatches manifest themselves through the transfer matrices.

## III. CALCULATION RESULTS

### 3.1 Delta barrier structure

Our model allows calculation of transport characteristics of a wide range of semiconductor-superconductor structures with various potentials. For the sake of simplicity, we have chosen a step-like spatial dependence of the superconductor order parameter $\Delta(x) = \Delta_0 \Theta(x)$, where $\Theta(x)$ is the Heaviside function.

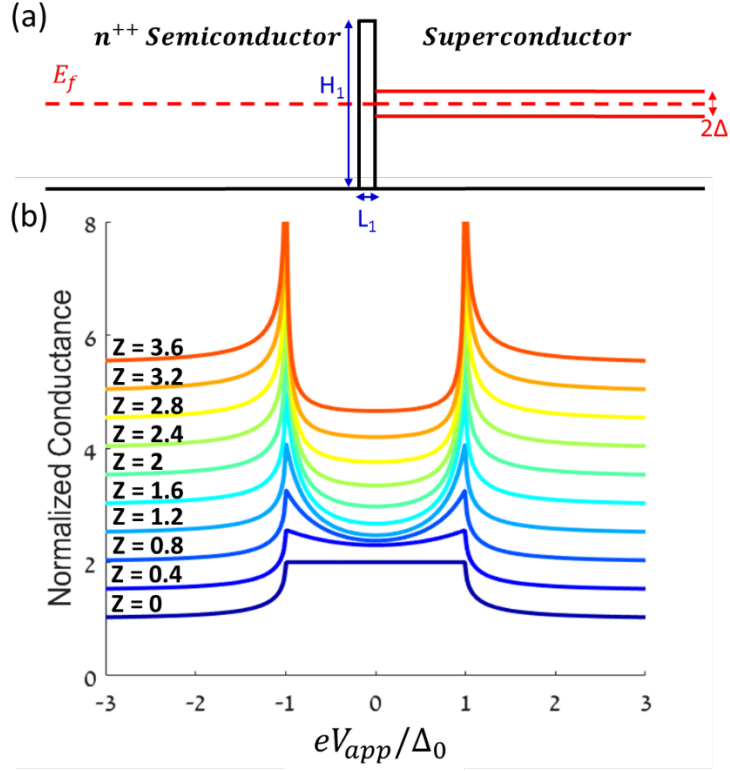

**Fig 2. (a)** A schematic drawing of the delta barrier structure. **(b)** Calculated results based on our model for different values of barrier strength Z, which coincide with the analytical results obtained used the BTK model. The curves are shifted vertically for clarity. For large values of Z, the tunneling process becomes dominant over the Andreev reflection process.

| **Delta Barrier** | |
|---|---|
| Potential $V(x)$ | $H_1[\Theta(x + L_1) - \Theta(x)]$ |
| Effective mass $m(x)$ | $m_1$ |
| Superconducting gap $\Delta(x)$ | $\Delta_1 \Theta(x)$ |
| Parameters | $0 \leq H_1 \leq 700\Delta_1, L_1 = 0.002S, m_1 = m_e, \Delta_1, S, L = 0.8S, N_1 = 5000, N_2 = 5000$ |

**Table 1.** Summary of the parameters used for the delta barrier structure. $\Theta(x)$ represents the Heaviside step function, $H_1$ and $L_1$ represent the height and width of the square barrier respectively, $\Delta_1$ represents the superconducting gap parameter, $m_1$ is the charge carrier effective mass, $m_e$ is the free electron mass, $S$ and $L$ are the pre-barrier and post barrier lengths, and $N_1$ and $N_2$ represent the number of segments in the barrier and termination regions respectively. In the limit of very small $L_1$ and large $H_1$, $H_1[\Theta(x+L_1)-\Theta(x)] \to H\delta(x)$, corresponding to a delta barrier with a strength parameter $H$.

In order to verify our model in comparison to the BTK model, we have applied it to the simple case of the delta barrier (Fig 2). the delta-potential barrier has been modeled as a high and narrow rectangular barrier where the parameter $H$ in $H\delta(x)$ is the product of the height and the width of the rectangular barrier. The $Z$ parameter is then defined as $Z=k_F H/2E_F$ where $k_F$ and $E_F$ are the Fermi wavenumber and energy respectively [15]. As the delta-potential barrier was modeled based on a rectangular barrier, its incorporation into the transfer matrices was similar to any other barrier, with the segments having to be sufficiently small due to the narrow width of the barrier. Figure 2 shows that the results obtained using our model match the analytical results provided by the BTK model.

### 3.2 Schottky and double barrier structures

Real devices, however, can feature more intricate, spatially varying potentials. A well-studied potential in non-superconducting structures is the Schottky barrier [34], which forms at metal-semiconductor junctions (Fig 3a) and is therefore important for modeling semiconductor-superconductor interfaces. Since the tunneling process becomes dominant with an increase of either the width or height of the Schottky barrier, strong doping is generally used at the vicinity of the junction in order to reduce the width of the Schottky barrier [34].

Our calculations provide a full description of superconductor-semiconductor interface with a Schottky barrier. Furthermore, we show that since the Schottky potential is asymmetric relative to the Fermi level, this results in an asymmetry in the conductance spectrum of the device (Fig 3b,c). This calculation provides theoretical modelling for the experimental results previously obtained [1,35]. It is worth noting that older experiments [35] often present results in the form of resistance and not conductance, so that e.g. a zero-bias peak in resistance corresponds to a zero-bias dip in differential conductance. For the case of asymmetry due to the presence of a Schottky potential barrier, our calculations show that the difference between the asymmetric peaks in the conduction spectrum amounts to ~3%. This is consistent with previous experimental results [35] which show varying degrees of asymmetry around 2-3%. A variety of material-interface and transport related effects were previously considered as possible reasons for the asymmetry, yet they were ruled out since they are completely symmetric in nature [35]. Non-ideality of retro-reflection of the electrons and holes, along with deviations from the spherical Fermi surface at the junction, were suggested to be considered as the reasons behind the asymmetry.

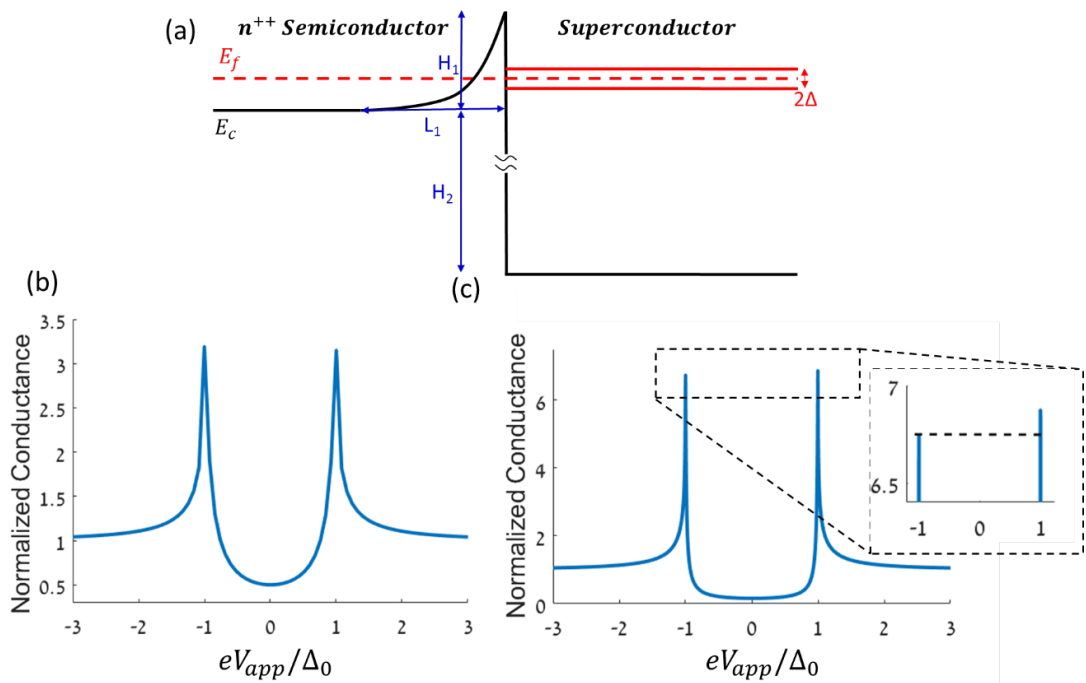

**Fig 3. (a)** A schematic drawing of a Schottky barrier between a semiconductor and a superconductor. The Fermi level lies above the heavily n-doped semiconductor conduction band edge $E_c$. **(b)** The Schottky barrier normalized conductance around the Fermi level **(c)** Example of an asymmetric conductance spectra caused by the asymmetric nature of a wide Schottky barrier. The inset shows the asymmetry of the peaks around 3%.

| Schottky barrier | |
|---|---|
| Potential $V(x)$ | $\left(\frac{H_1}{L_1^2}x^2 + \frac{2H_1}{L_1}x + H_1\right)[\Theta(x+L_1) - \Theta(x)] + H_2\Theta(-x)$ |
| Effective mass $m(x)$ | $m_1\Theta(-x) + m_2\Theta(x)$ |
| Superconducting gap $\Delta(x)$ | $\Delta_1\Theta(x)$ |
| Parameters | $H_1 = 100\Delta_1, H_2 = 2600\Delta_1, L_1 = S, m_1 = 0.041 m_e, m_2 = m_e, \Delta_1, S, L = 3S/7, N_1 = 5000, N_2 = 5000$ |

**Table 2.** Summary of the parameters used for the Schottky barrier structure. $H_1$ represents the maximum height of the Schottky barrier, $L_1$ represents the total length of the Schottky barrier, $H_2$ represents the potential offset between the different materials on both sides of the junction, $m_1$ and $m_2$ are the effective mass on each side of the junction. The shape of the Schottky barrier is parabolic assuming the depletion region approximation [36].

Our model takes into account retro-reflection non-idealities which arise as the result of the barrier present in the problem as well as Fermi velocity mismatches due to different effective masses and *k*-vectors

  Furthermore, we show that it is possible to significantly enhance Andreev reflection and thus Cooper-pair injection into the semiconductor, by the use of resonant tunneling through a specially designed double barrier. In this method, a barrier structure can be designed such that resonant energy levels exist in the barrier, with the simplest form being a double barrier. Resonant energy levels have been shown to enable full transmission without reflection of charge carriers [37]. Our calculations show that a Fermi level aligned with one of the resonant levels provides strong enhancement of the Andreev reflection process, with conductance reaching twice the normal-to-normal conductance. However, misalignment of the Fermi level with any of the resonant levels can cause strong suppression of the Andreev reflection process in favor of the regular, non-resonating tunneling – similar to the single-barrier case.

  We propose a practically feasible double barrier obtained by adding a square potential barrier to the Schottky potential (Fig 4 a). Such a barrier can be designed by using a semiconductor heterostructure for the square potential and a semiconductor-superconductor interface for the Schottky potential, with enhanced Andreev reflection obtained at resonant energy levels. While the resonant effect we observe is caused solely due to the shape of the barrier, it was previously shown that reflection of quasiparticles between a superconducting gap and an offset delta potential [38] can also result in resonant effects. The various conduction spectra obtained from the double barrier potential lead to two important observations.

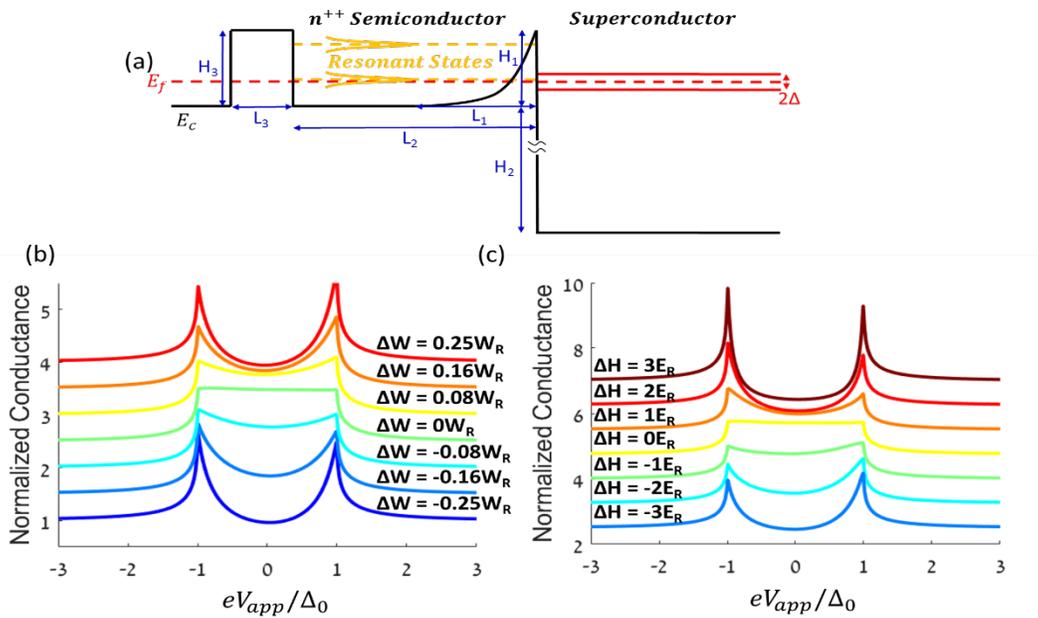

**Fig 4. (a)** A schematic drawing of a Schottky barrier with an additional square barrier used to form resonant energy levels (yellow). Alignment of the Fermi level (red) with a resonant energy level results in a strong enhancement of the Andreev reflection process. **(b)** Normalized conductance for various well width offsets where $\Delta W = 0$ where resonant tunneling is achieved. $W_R$ is the width of the well that results in bound state energy coinciding with $E_f$. **(c)** Normalized conductance for various square barrier height offsets where $\Delta H = 0$ is the value for which resonant tunneling is achieved. $E_R$ is the value of the resonant energy level relative to the bottom of the barrier. For both (b) and (c), a suppression of the Andreev reflection process can be observed, confirming the ability of resonant energy states to enhance Andreev reflection. The curves are shifted vertically for clarity.

| Schottky + Square barrier | |
|---|---|
| Potential $V(x)$ | $\left(\frac{H_1}{L_1^2}x^2 + \frac{2H_1}{L_1}x + H_1\right)[\Theta(x+L_1) - \Theta(x)] + H_2\Theta(-x) + H_3[\Theta(x+L_2+L_3) - \Theta(x+L_2)]$ |
| Effective mass $m(x)$ | $m_1\Theta(-x) + m_2\Theta(x)$ |
| Superconducting gap $\Delta(x)$ | $\Delta_1\Theta(x)$ |
| Parameters | $H_1 = 200\Delta_1, H_2 = 2600\Delta_1, 0 \leq H_3 \leq 400\Delta_1, L_1 = 0.6S, L_2 = 0.8S, L_3 = 0.1S, m_1 = m_e, m_2 = m_e, \Delta_1, S, L = 0.8S, N_1 = 5000, N_2 = 5000$ |

**Table 3.** Summary of the parameters used for the Schottky double barrier structure. $H_3$ represents the height of the square barrier, $L_2$ represents the distance of the square barrier from the junction and $L_3$ represent its width. Since a barrier structure can be obtained by materials with very similar properties (e.g. alloys of $Al_xGa_{1-x}As$ with varying x), the effective mass in the barrier was assumed to be equal to that of the rest of semiconductor side.

The first observation is that although Schottky barriers are limiting factors when designing proper contacts, they can be utilized to form double barriers containing resonant levels. These in turn can be used to enhance transmission, thus giving Schottky barriers a useful role in the device. The second important observation relates to the factors required to achieve enhancement of the Andreev reflection process. While reducing the difference $\Delta E = E_f - E_{res}$ (with the resonant level centered at $E_{res}$) is a key requirement for alignment, the superconducting gap parameter $\Delta$ and the characteristic width of the resonant level $\Gamma$, must also be taken into account (Fig 5), with three important regimes to be considered.

$\Delta \gg \Gamma$ regime: The first regime is for the case of $\Delta \gg \Gamma$. In this regime, the width of the resonant level is much smaller than the superconducting gap. This has an important implication: while enhanced Andreev reflection does occur, it only occurs in a small energy range of the superconducting gap determined by the resonant tunneling bandwidth $\Gamma$ rather than occurring all across the superconducting gap Δ. Moreover, since the resonant level has a characteristic shape, the conductance spectrum assumes the shape of the resonant level rather than the flat shape which is characteristic of strong Andreev reflection.

Δ~$\Gamma$ regime: The second regime is for Δ~$\Gamma$, where the width of the resonant level is on the order of the width of the superconducting gap. This results in the conductance spectrum shape being effected by both the characteristic shape of Andreev reflection and the shape of the resonant level. Moreover, changes to the energy difference Δ$E$ have a strong effect on Andreev reflection being enhanced or suppressed as the overlap between Δ and $\Gamma$ becomes difficult to achieve.

Δ≪ $\Gamma$ regime: The third regime is for the relation Δ≪ $\Gamma$. In this regime, the width of the resonant level is much larger than that of the superconducting gap. This results in the conductance spectrum assuming the characteristic shape of the Andreev reflection spectrum. This regime is the preferred one, since a broad resonance relaxes the requirement on Fermi level alignment with the resonant level. Figures 4b and 4c demonstrate the importance of this requirement relaxation as small changes to the width and height of the barrier cause strong suppression of the Andreev reflection process. Moreover, through proper design of the barrier, it is possible to control both $E_{res}$ and $\Gamma$ with the ability to increase or decrease the magnitude of either parameter. Such control is not possible for Δ as it strongly depends on the superconductor used in the device as well as the temperature, and can only be increased by either further cooling the device or replacing the superconductor used, with both options being much less feasible than proper barrier design. Figure 5 demonstrates a sweep of Δ through all three regimes, with each regime having a conductance spectrum that exhibits the magnitude relation between Δ and $\Gamma$. One of the notable features is the strong presence as well as asymmetry of side lobes in the third domain (Δ≫ $\Gamma$). The side lobes are a characteristic of the tunneling process (coherence peaks) which appears when Andreev reflection is suppressed. The conductance spectrum (Fig. 5) exhibits both characteristics of Andreev reflection (at the center) and tunneling (at the side lobes) as a result of leaving the range of the resonant energy level, which causes a shift from one process to another.

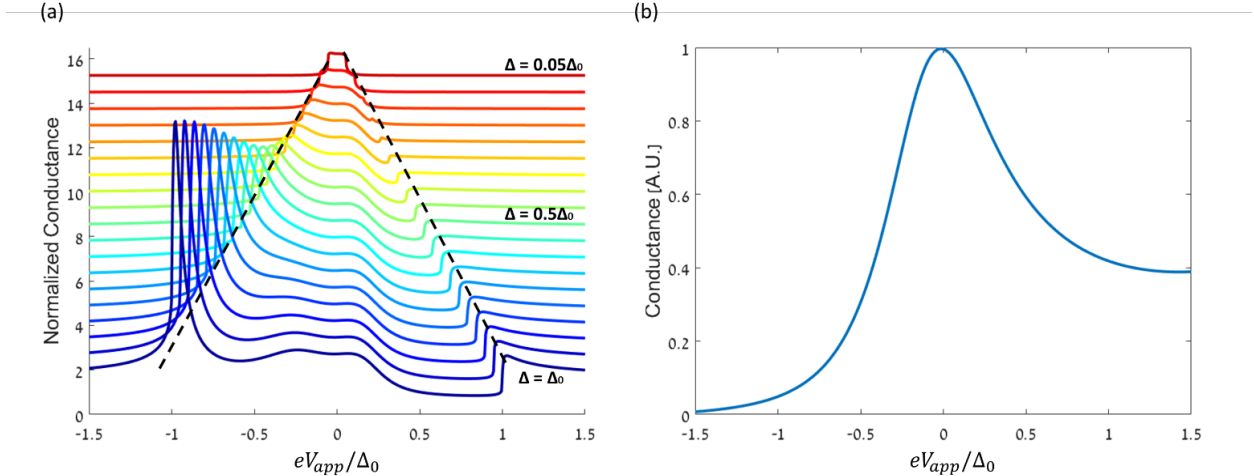

**Fig 5**. **(a)** $\Delta$ sweep showing all three regimes for the $\Delta$ and $\Gamma$ interplay with $\Gamma \sim \Delta_0$. The curves are shifted vertically for clarity. For $\Delta \gg \Gamma$ (lower third of graph), the conductance spectrum obtains the shape of the resonant energy level. For $\Delta \sim \Gamma$ (middle third of graph), the conductance spectrum obtains a shape affected by both the resonant energy level and Andreev reflection. For $\Delta \ll \Gamma$ (upper third of graph), the conductance spectrum obtains the characteristic shape of Andreev reflection. The curves are shifted vertically for clarity. **(b)** Electron transmission spectrum without superconducting gap.

The strong asymmetry of the side lobes stems from the asymmetry of the resonant energy level (Fig 5b). Higher conductance is expected at higher energies as the effective barrier seen by the particles becomes smaller. This is evident from the graph as the right lobe is much smaller than the left lobe, indicating a stronger contribution from the tunneling process (compared to the Andreev reflection process) on the left side.

## 3.3 Superlattice barrier structure

The idea behind the double barrier resonant states can be expanded to the case of multiple barriers – a superlattice [39] (Fig 6a). In such superlattices, resonant bands exist instead of discrete resonant energy levels. The typical width of such bands can be designed to be much larger than Δ, resulting in relaxation of the requirement to align the Fermi level with the resonance.

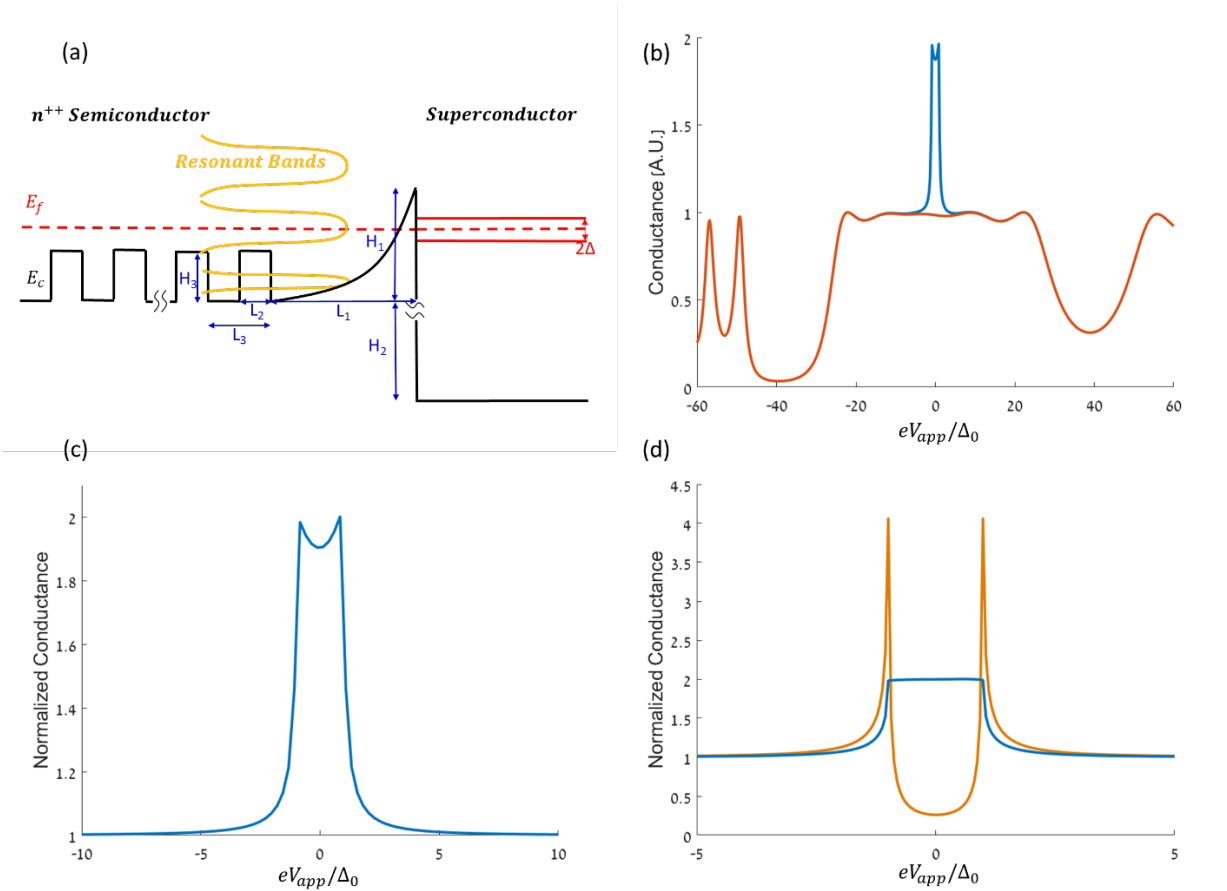

**Fig 6. (a)** A schematic drawing of the barrier structure for the case of a superlattice with resonant bands (yellow) and a Schottky barrier. **(b)** Electron transmission coefficient for the case of 4 square barrier cells with (blue line) and without (orange line) superconducting gap. **(c)** Normalized conductance with the Fermi level inside one of the bands. **(d)** Conductance spectrum with (blue) and without (orange) a superlattice structure.

| **Superlattice + Schottky barrier** | |
|---|---|
| Potential $V(x)$ | $\sum_{d=1}^{D} H_3[\Theta(x + L_2 + dL_3) - \Theta(x + dL_3)]$ $+ \left(\frac{H_1}{L_1^2}x^2 + \frac{2H_1}{L_1}x + H_1\right)[\Theta(x + L_1) - \Theta(x)] + H_2\Theta(-x)$ |
| Effective mass $m(x)$ | $m_1\Theta(-x) + m_2\Theta(x)$ |
| Superconducting gap $\Delta(x)$ | $\Delta_1\Theta(x)$ |
| Parameters | $H_1 = 100\Delta_1, H_2 = 2600\Delta_1, H_3 = 100\Delta_1, L_1 = 0.1S, L_2 = 0.1S, L_3 = 0.2S, D = 4, m_1 = m_e, m_2 = m_e, \Delta_1, S, L = 0.8S, N_1 = 5000, N_2 = 5000$ |

**Table 4.** Summary of the parameters used for the superlattice barrier structure. $H_3$ represents the height of each square barrier, $L_2$ is the width of each square barrier, $L_3$ is the distance between two such barriers and $D$ is the total number of square barriers in the superlattice.

Multiple barriers form resonant energy bands (Fig 6b) which allow a much stronger Andreev reflection (Fig 6c) to occur. Moreover, the width of the bands increases with increasing quasiparticle energy, while the forbidden gaps between the bands decrease. Control over the number of bands and their distribution is possible through engineering of the superlattice.

In order to demonstrate that the enhancement of Andreev reflection results from the presence of the superlattice, we calculated the conductance spectrum with and without the superlattice (Fig 6d). It is possible to note the strong difference the presence of a superlattice has on the resulting conductance spectrum, ranging from strong suppression of the Andreev reflection process to strong enhancement (to the maximum).

## IV. CONLUSIONS

We have shown that our model provides new insights into the behavior of nanoscale interfaces in hybrid semiconductor-superconductor structures such as asymmetrical conductance spectra occurring due to the inherent asymmetry of the barriers relative to the Fermi level. Such asymmetric shapes have been observed in various experiments, and although some hypotheses have been put forward previously for the origin of the asymmetry [1, 33], no theoretical modelling has been performed so far. We have shown, both through theory and numerical calculations, that non-ideality of the quasiparticle retro-reflection due to the barrier structure can result in asymmetry. In addition, we have shown that the Schottky barrier, which is generally regarded as a significant limiting issue when designing proper contacts and junctions, could have a strong contribution to the conduction spectrum when combined with additional features in the barrier. The double barrier is an example of a structure which makes use of the Schottky barrier at the junction to form resonant energy levels which can strongly enhance the Andreev reflection process. Moreover, we have shown that the concept of the double barrier can be expanded to superlattice structures which greatly enhance Andreev reflection while simultaneously relaxing the condition for matching the Fermi level to resonance. Our results have great potential to increase the overall efficiency of superconductor-semiconductor devices through enhanced Andreev reflection which corresponds to much more efficient Cooper-pair injection.

**Acknowledgement**

This research was supported by the ISF-NSFC joint research program (grant No. 2220/5) and ISF FIRST Program (grant No. 1995/17).

**REFERENCES**

[1] A. Kastalsky, A. W. Kleinsasser, L. H. Greene, R. Bhat, F. P. Milliken and J. P. Harbison, "Observation of pair currents in superconductor-semiconductor contacts", Phys. Rev. Lett. **67**, 3026 (1991).

[2] C. Nguyen, H. Kroemer and E. L. Hu, "Anomalous Andreev conductance in InAs-AlSb quantum well structures with Nb electrodes", Phys. Rev. Lett. **69**, 2847 (1992).

[3] T. Nishino, M. Hatano, H. Hasegawa, T. Kure and F. Murai, "Carrier reflection at the superconductor-semiconductor boundary observed using a coplanar-point-contact injector", Phys. Rev. B **41**, 7274(R) (1990).

[4] D. R. Heslinga, S. E. Shafranjuk, H. van Kempen and T. M. Klapwijk, "Observation of double-gap-edge Andreev reflection at Si/Nb interfaces by point-contact spectroscopy", Phys. Rev. B **49**, 10484 (1994).

[5] S. S. Mou, H. Irie, Y. Asano, K. Akahane, H. Kurosawa, H. Nakajima, H. Kumano, M. Sasaki and I. Suemune, "Superconducting Light-Emitting Diodes", IEEE J. Sel. Top. Quantum Electron, **21**, 7900111 (2015).

[6] S. S. Mou, H. Irie, Y. Asano, K. Akahane, H. Nakajima, H. Kumano, M. Sasaki, A. Murayama, and I. Suemune, "Optical observation of superconducting density of states in luminescence spectra of InAs quantum dots", Phys. Rev. B. **92**, 035308 (2015).

[7] I. Suemune, T Akazaki, K. Tanaka, M. Jo, K. Uesugi, M. Endo, H. Kumano, E. Hanamura, H. Takayanagi, M. Yamanishi and H. Kan, "Superconductor-Based Quantum-Dot Light-Emitting Diodes: Role of Cooper Pairs in Generating Entangled Photon Pairs", Jpn. J. Appl. Phys. **45**, 9264 (2006).

[8] M. Khoshnegar and A. H. Majedi, "Entangled photon pair generation in hybrid superconductor–semiconductor quantum dot devices", Phys. Rev. B **84**, 104504 (2011).

[9] I. Suemune, Y. Hayashi, S. Kuramitsu, K. Tanaka, T. Akazaki, H. Sasakura, R. Inoue, H. Takayanagi, Y. Asano, E. Hanamura, S. Odashima, and H. Kumano, "A Cooper-Pair Light-Emitting Diode: Temperature Dependence of Both Quantum Efficiency and Radiative Recombination Lifetime", Appl. Phys. Express **3**, 054001 (2010).

[10] E. Sabag, S. Bouscher, R. Marjieh, and A. Hayat, "Photonic Bell-State Analysis Based on Semiconductor-Superconductor Structures ", Phys. Rev. B **95**, 094503 (2017).

[11] R. Marjieh, E. Sabag and A. Hayat, "Light amplification in semiconductor-superconductor structures", New J. Phys. **18**, 023019 (2016).

[12] A. F. Andreev, "Thermal conductivity of the intermediate state of superconductors". Sov. Phys. JETP. **19**, 1228 (1964).

[13] Y. Asano, I. Suemune, H. Takayanagi, and E. Hanamura, "Luminescence of a Cooper Pair", Phys. Rev. Lett. **107**, 073901 (2011).

[14] A. Hayat, H.Y. Kee, K. S. Burch and A. M. Steinberg, "Cooper-pair-based photon entanglement without isolated emitters", Phys. Rev. B. **89**, 094508 (2014).

[15] G.E Blonder, M. Tinkham, and T.M. Klapwijk, "Transition from metallic to tunneling regimes in superconducting microconstrictions: Excess current, charge imbalance and supercurrent conversion", Phys. Rev. B. **25**, 4515 (1982).

[16] R. P. Panguluri, K. C. Ku, T. Wojtowicz, X. Liu, J. K. Furdyna, Y. B. Lyanda-Geller, N. Samarth and B. Nadgorny, "Andreev reflection and pair-breaking effects at the superconductor/magnetic semiconductor interface", Phys. Rev. B. **72**, 054510 (2005).

[17] T. Yamashita, H. Imamura, S. Takahashi and S. Maekawa, "Andreev reflection in ferromagnet/superconductor/ferromagnet double junction systems", Phys. Rev. B. **67**, 094515 (2003).

[18] D. Beckmann and H. B. Weber, "Evidence for Crossed Andreev Reflection in Superconductor-Ferromagnet Hybrid Structures", Phys. Rev. Lett. **93**, 197003 (2004).

[19] Y. De Wilde, J. Heil, A. G. M. Jansen, P. Wyder, R. Deltour, W. Assmus, A. Menovsky, W. Sun, and L. Taillefer, "Andreev reflections on heavy-fermion superconductors", Phys. Rev. Lett. **72**, 2278 (1994).

[20] W. K. Park, J. L. Sarrao, J. D. Thompson, and L. H. Greene, "Andreev Reflection in Heavy-Fermion Superconductors and Order Parameter Symmetry in $CeCoIn_5$", Phys. Rev. Lett. **100**, 177001 (2008).

[21] N. A. Mortensen, K. Flensberg, and A.P. Jauho, "Angle dependence of Andreev scattering at semiconductor–superconductor interfaces", Phys. Rev. B. **59**, 10176 (1999).

[22] S. Kashiwaya, Y. Tanaka, M. Koyanagi and K. Kajimura, "Theory for tunneling spectroscopy of anisotropic superconductors", Phys. Rev. B **53**, 2667 (1996).

[23] J. Bardeen, L. N. Cooper, and J. R. Schrieffer, "Microscopic Theory of Superconductivity", Phys. Rev. **106**, 162 (1957).

[24] C. Weisbuch and B. Vinter, "Quantum Semiconductor Structures – Fundamentals and Applications", Academic Press (1991).

[25] E. Burstein and S. Lundqvist, "Tunneling Phenomena in Solids", Plenum Press (1969).

[26] Kuei Sun and Nayana Shah, " General framework for transport in spin-orbit-coupled superconducting heterostructures: Nonuniform spin-orbit coupling and spin-orbit-active interfaces", Phys. Rev. B **91**, 144508 (2015).

[27] Jean-Marc Lévy-Leblond, "Position-dependent effective mass and Galilean invariance", Phys. Rev. A **52**, 1845 (1995).

[28] P. Monk, "Finite Element Methods for Maxwell's Equations", 1st edition, Oxford Science Publications (2003).

[29] J. D. Jackson, "Classical Electrodynamics", 3rd edition, John Wiley & Sons (1998).

[30] B. Jonsson and S.T. Eng, " Solving the Schrodinger Equation in Arbitrary Quantum-Well Potential Profiles Using the Transfer Matrix Method", IEEE Journal of Quantum Electronics, **26**, 11 (1990).

[31] C. Jirauschek, "Accuracy of Transfer Matrix Approaches for Solving the Effective Mass Schrödinger Equation", IEEE Journal of Quantum Electronics, **45**, 9 (2009).

[32] L. Esaki and R. Tsu, "Superlattice and Negative Differential Conductivity in Semiconductors", IBM Journal of Research and Development **14**, 1 (1970).

[33] R. Pérez-Alvarez and H. Rodriguez-Coppola, "Transfer Matrix in 1D Schrödinger Problems with Constant and Position-Dependent Mass", Phys. Status Solidi, **145**, 493 (1988).

[34] R. F. Pierret, "Semiconductor Device Fundamentals", Addison Wesley, 2nd edition (1996).

[35] H. F. C. Hoevers, M. G. D. van der Grinten, P. L. H. Jennen, H. van Kempen and P. C. van Son, "Asymmetric differential resistance of point contacts on normal-metal-superconductor bilayers", J. Phys. Cond. Matter **6**, 65 (1994).

[36] Y. Taur and T. H. Ning, "Fundamentals of Modern VLSI Devices", 2nd edition, Cambridge University Press (2009).

[37] L. L. Chang, L. Esaki and R. Tsu, "Resonant tunneling in semiconductor double barriers", Appl. Phys. Lett. **24**, 593 (1974).

[38] P. C. van Son, H. van Kempen and P. Wyder, "Andreev reflection and geometrical resonance effects for a gradual variation of the pair potential near the normal-metal/superconductor interface", Phys. Rev. B. **37**, 5015 (1988).

[39] R. Tsu and L. Esaki, "Tunneling in a finite superlattice", Appl. Phys. Lett. **22**, 562 (1973).